\documentclass[preprint,authoryear]{elsarticle}

\usepackage{graphicx,subfigure}
\usepackage{epstopdf}
\usepackage{amsmath, amssymb}
\usepackage{rotating}
\usepackage{morefloats}
\usepackage{comment}
\usepackage{natbib}
\usepackage{caption}

\begin{document}
\title{Maximum drawdown, recovery, and momentum}
\author[sb]{Jaehyung Choi\corref{cor1}}
\ead{jj.jaehyung.choi@gmail.com}

\cortext[cor1]{Correspondence address: Goldman Sachs \& Co., New York, NY 10282, USA.}
\address[sb]{Goldman Sachs \& Co., New York, USA
\footnote{Disclosure: The opinions and statements expressed in this paper are those of the author and may be different to views or opinions otherwise held or expressed by or within Goldman Sachs. The content of this paper is for information purposes only and is not investment advice or advice of any other kind. None of the author, Goldman Sachs, or its affiliates, officers, employees, or representatives accepts any liability whatsoever in connection with any of the content of this paper or for any action or inaction of any person taken in reliance upon such content or any part thereof.}}

\begin{abstract}
	We empirically test predictability on asset price by using stock selection rules based on maximum drawdown and its consecutive recovery. In various equity markets, monthly momentum- and weekly contrarian-style portfolios constructed from these alternative selection criteria are superior not only in forecasting directions of asset prices but also in capturing cross-sectional return differentials. In monthly periods, the alternative portfolios ranked by maximum drawdown measures exhibit outperformance over other alternative momentum portfolios including traditional cumulative return-based momentum portfolios. In weekly time scales, recovery-related stock selection rules are the best ranking criteria for detecting mean-reversion. For the alternative portfolios and their ranking baskets, improved risk profiles in various reward-risk measures also imply more consistent prediction on the direction of assets in future. In the Carhart four-factor analysis, higher factor-neutral intercepts for the alternative strategies are another evidence for the robust prediction by the alternative stock selection rules.
\end{abstract}


\maketitle
\section{Introduction}
	Seeking statistical arbitrages in financial markets is the most important task to both academics and practitioners in finance. Empirically-existent systematic arbitrages are not only experimental counter-examples to the efficient market hypothesis proposed by \cite{Fama:1965p4897} and \cite{Samuelson:1965p4904}, but also the sources of lucrative trading strategies to the practitioners. Among various market inefficiencies, price momentum (\cite{Jegadeesh:1993p200}) is one of the most well-known market anomalies that are not explained by the Fama-French three-factor model (\cite{Fama:1993, Fama:1996}). Although the price momentum has been found in many asset classes and markets (\cite{Rouwenhorst:1998,Rouwenhorst:1999,Okunev:2003,Erb:2006,Asness:2013,Moskowitz:2010}), it is still not fully explicable with numerous alternative explanations on the anomaly such as autocorrelation and cross-sectional correlation (\cite{Lo:1990p883, Lewellen:2002}), sector momentum (\cite{Moskowitz:1999p4294}), investors' behavioral aspects to news dissemination (\cite{Hong:1999p4506, Terence:1998p4385, Daniel:1998p4514, Barberis:1998p307}), symmetry breaking of arbitrage (\cite{Choi:2012}), and transaction cost (\cite{Lesmond:2004}).
	
	Introducing various ranking rules to momentum-style portfolio construction is a worthwhile approach for understanding the price momentum. First of all, it is a direct way of searching for potential factors which can explain the momentum phenomena. Additionally, such alternative criteria can be used as potential indicators for trading signals in practice. Several selection rules have been proposed and the literature covers time series model (\cite{Moskowitz:2010}), trading volume (\cite{Lee:2000}), liquidity (\cite{ Amihud:1986, Hu:1997,Datar:1998, Amihud:2002, Kim:2012}), analogy from momentum in physics (\cite{Choi:2012a}), reward-risk measures (\cite{Rachev:2007p616,Choi:2015}), and 52-week high price (\cite{George:2004, Liu:2011, George:2018}). 
	
	In particular, the 52-week high momentum (\cite{George:2004, Liu:2011, George:2018}) is in line with the philosophy of price momentum. Since the 52-week high price is the highest price during recent 52 weeks, higher returns to the highest price in the most recent one year horizon underline the existence of stronger price momentum during the period. Achieving the new 52-week high price is also attracting investors' attentions and interests based on the positive aspiration that assets are capable of maintaining upward price movements in following periods. Although consideration on the 52-week low price is also plausible in the similar analogy, the opposite direction has not been reported as a successful ranking criterion for the momentum strategy.
	
	Testing the 52-week low price as a selection rule is not the only approach for incorporating downside risks into the momentum-style portfolio construction process. \cite{Rachev:2007p616} paid attention to reward-risk measures such as Sharpe ratio, STAR ratio, and Rachev ratio. It was reported that the tail behavior and risk can predict future price directions of equities in the S\&P 500 component universe. It is also noteworthy that their alternative portfolios were less riskier with thinner downside tails. As an extension of the work by \cite{Rachev:2007p616}, reward-risk momentum strategies using classical tempered stable (CTS) distributions were implemented in various asset classes (\cite{Choi:2015}). Regardless of asset class and market, it was found that the alternative strategies ranked by the CTS reward-risk measures exhibit improved profitability and reward-risk profiles. Similarly, diversified reward-risk parity (\cite{Choi:2021}) that penalizes riskier assets and rewards less riskier assets in portfolio construction also delivers more advanced performance and risk management under less complicated portfolio construction methods. These results are also consistent with the low-volatility anomaly (\cite{Blitz:2007, Baker:2011}).
	
	One of the most popular risk measures in practice is maximum drawdown defined as the worst cumulative loss in a given period. Maximum drawdown is also used in the definitions of the Calmar ratio and the Sterling ratio in order to assess the performance and risk of mutual funds and hedge funds. Several advantages of maximum drawdown over Value-at-Risk (VaR) and conditional VaR (CVaR) are the followings. First of all, it is more insightful than VaR and CVaR. When two historical price charts are given, it is more straightforward to recognize which asset has the smaller maximum drawdown. Additionally, since it is computed from a simple sum of log-returns during the drawdown period, it is easier to calculate maximum drawdown directly from time series. Moreover, the maximum drawdown is a model-free risk measure. For example, VaRs and CVaRs calculated from Gaussian distributions are different with the risk measures from heavy-tailed distributions. Finally, maximum drawdown encodes much information on asset price evolution in time. Even though parameters of a given probability distribution are identical, a random permutation of a given time series produces a different maximum drawdown.

	In this paper, we introduce multiple composite ranking criteria originated from maximum drawdown and successive recovery. By adopting these ranking rules, we construct alternative momentum/contrarian-style portfolios in order to test predictability of the alternative ranking measures. Monthly momentum and weekly contrarian strategies based on the alternative stock selection rules are implemented in U.S. and South Korea stock markets. It is noteworthy that the new selection rules exhibit enhanced predictability on asset price such that the alternative strategies outperform the traditional cumulative return-based strategies in performance and risk management. In particular, drawdown measures provide more consistent trend-following strategies in a monthly scale, and recovery criteria robustly work with the weekly strategies. The dominance in predictability and performance is also found in each ranking portfolio and the regression from the Carhart four-factor model. The paper is structured as follows. In next section, new ranking rules based on maximum drawdown and sequential recovery are defined. In section \ref{sec_mddr_momentum_data_method}, datasets and methodologies are introduced. Performance and reward-risk profiles of the alternative portfolios are presented in section \ref{sec_mddr_momentum_result}. Factor analysis on momentum returns is given in section \ref{sec_mddr_momentum_factor_analysis}. We conclude the paper in section \ref{sec_mddr_momentum_conclusion}.

\section{Construction of alternative stock selection rules} 
\label{sec_mddr_momentum_selection_rule}
	As mentioned before, maximum drawdown is the worst loss among successive declines from peaks to troughs during a given period. At time $T$, it is defined as
	\begin{equation}
		\mathrm{MDD}=\overset{}{\underset{\tau\in(0,T)}{\textrm{max}}}\Big(\overset{}{\underset{t\in(0,\tau)}{\textrm{max}}} \big(P(t)-P(\tau)\big)\Big)\nonumber
	\end{equation}
	where $P(t)$ is the log-price at time $t$. It can be expressed with log-returns:
	\begin{equation}
		\mathrm{MDD}=-\overset{}{\underset{\tau\in(0,T)}{\textrm{min}}}\Big(\overset{}{\underset{t\in(0,\tau)}{\textrm{min}}} R(t,\tau)\Big)\nonumber
	\end{equation}
	where $R(t,\tau)$ is the log-return between $t$ and $\tau$. The maximum drawdown is regarded the worst-case scenario to an investor who starts his/her investment in the period. It is obvious for investors to prefer smaller maximum drawdowns to larger drawdowns in portfolio performance.
	
	It is noteworthy that maximum drawdown is closely related to price momentum. For example, the direction and magnitude of price momentum are affected by maximum drawdown. For positive momentum, maximum drawdown is regarded a part of mean-reversion process, if the extent of the maximum drawdown is small. Meanwhile, if it is large enough, the maximum drawdown is more likely to break the original upward trend and generates a new downward trend. Additionally, it is straightforward that the drawdown also contributes to downside momentum. Larger maximum drawdowns would be much preferred for short-selling assets.
	
	The successive recovery after the maximum drawdown is defined as
	\begin{equation}
		\mathrm{R}=R(t^{*},T)\nonumber
	\end{equation}
	where $t^{*}$ is the time moment at the end of the maximum drawdown formation. It imposes how much loss from the worst decline is recovered by the short-term reversion. Similar to maximum drawdown, it is also helpful to understand the price momentum. When an asset price is on the upward trend, it is regarded the support to the initial trend. Meanwhile, the recovery in downside momentum is a reversion against the overall trend. It is obvious that an asset with stronger recovery is more favored for long positions than an asset with weaker recovery. Contrary to the long positions, smaller recovery would be preferred for short-selling.
		
	In this regard, maximum drawdown and consecutive recovery are indispensable in the processes of detecting and analyzing signals for price trending and its reversion. It is obvious that these aspects need to be incorporated into momentum analysis and portfolio construction. Additionally, it is possible to develop new composite selection rules stemming from the maximum drawdown and the recovery. For example, even if two assets have the same cumulative return, some investors would penalize assets with worse maximum drawdowns in portfolio construction based on investors' investment goals and risk appetites. In another case, an asset with stronger short-term recovery would be preferred in the alternative ranking systems. The traditional ranking criterion is impossible to discern the characteristics of the price evolution in past history.
	
	Before developing new alternative selection rules, we need to take a closer look on cumulative return. For a given asset, the time window of asset price evolution is chronologically decomposed into the following three sub-periods. The first sub-period is the peak formulation phase that ranges from the beginning of the estimation period to the peak prior to the maximum downfall. In the next sub-period, the maximum drawdown formation has progressed. The last sub-period is the recovery stage from the completion moment of the maximum drawdown formulation to the end of the estimation period. The above decomposition on the cumulative log-return $\mathrm{C}$ is represented with the log-returns from these three different phases:
	\begin{align}
		\mathrm{C}&=R_{I}+R_{II}+R_{III}\nonumber\\
		&=\mathrm{PP}-\mathrm{MDD}+\mathrm{R}\nonumber
	\end{align}
	where PP, MDD, and R are the log-returns during the pre-peak sub-period, drawdown sub-period, and recovery sub-period, respectively.
	
	Exploiting the above decomposition, it is possible to construct hybrid indicators based on cumulative return, maximum drawdown, and recovery. Taking a weighted average with extra weights on certain specific sub-periods is one way of developing new stock selection or ranking rules. Possible combinations with maximum drawdown and consecutive recovery are given in Table \ref{tbl_mddr_ranking_criteria_desc}. Since the pre-peak return $\mathrm{PP}$ is based on relatively older information on price evolution, it could be erased by maximum drawdown and adjacent recovery, and this is why we exclude alternative rules related to $\mathrm{PP}$.
	
\begin{table}[h!]
\begin{center}
\caption{Description on alternative selection rules using maximum drawdown and recovery}
\small
\begin{tabular}{c l c}
\hline
Portfolio name & Criterion & Weights on $R_{t}$\\
\hline 
C & Cumulative return& (1,1,1)\\
M & MDD & (0,1,0)\\
R & Recovery & (0,0,1)\\
RM & Recovery-MDD & (0,1,1)\\
CM & Cumulative return-MDD & (1,2,1)\\
CR& Cumulative return+Recovery & (1,1,2)\\
CMR& Cumulative return-MDD+Recovery &(1,2,2)\\
\hline
\end{tabular}
\label{tbl_mddr_ranking_criteria_desc}
\end{center}
\end{table}

	Let us re-iterate meanings of alternative selection rules in Table \ref{tbl_mddr_ranking_criteria_desc}. The C strategy is considered as the benchmark strategy. It is the traditional momentum/contrarian strategy that employs the cumulative return in an estimation period as a ranking criterion. The M portfolio and the R portfolio are constructed from maximum drawdown and recovery, respectively. Additionally, more complicated selection rules are produced by different weights on certain periods. The RM strategy utilizes the difference between recovery and maximum drawdown. It indicates the net profit or loss such that the maximum drawdown is restored by the short-term recovery adjacent to the maximum drawdown. The CM strategy not only considers cumulative return but also simultaneously penalizes the maximum drawdown during the estimation period. This ranking criterion gives a penalty to assets with worse drawdowns under the same cumulative return. In the similar ways, it is possible to assign weights on only the recovery period or both the maximum drawdown sub-period and the recovery sub-period in order to construct the CR strategy and the CMR strategy, respectively.
	
\section{Dataset and methodology}
\label{sec_mddr_momentum_data_method}
\subsection{Dataset}
	The dataset for this study consists of KOSPI 200 universe, SPDR U.S. sector ETFs, and S\&P 500 universe.
\subsubsection{South Korea equity market: KOSPI 200}
	KOSPI 200 is a stock benchmark index that is the value-weighted and sector-diversified index of main 200 stocks in South Korea stock market. Historical price information and the component change list are downloaded from Korea Exchange. The period from January 2003 to December 2012 is considered.
\subsubsection{U.S. equity market: SPDR U.S. sector ETFs}
	Sector ETFs are selected for simulating sector momentum which adopts alternative stock selection rules. In particular, SPDR U.S. sector ETFs are chosen because it is the sector ETF collection in which the equal length of price history is available for all the sectors. The SPDR U.S. sector ETF universe includes XLB (Materials), XLE (Energy), XLF (Financial), XLI (Industrial), XLK (Technology), XLP (Consumer Staples), XLU (Utilities), XLV (Health Care), and XLY (Consumer Discretionary). The time span covers from January 1999 to December 2012, and the data source is Bloomberg.
\subsubsection{U.S. equity market: S\&P 500}
	The price information and the list of S\&P 500 historical component changes are collected from Bloomberg. The time window for the dataset starts from January 1993 and ends in December 2012.
\subsection{Portfolio construction processes}
	The basic methodology for portfolio construction is building momentum (or contrarian) style portfolios as given in \cite{Jegadeesh:1993p200}. Assets in a market universe are sorted by a given selection rule during an estimation period. The length of the estimation period is six months for momentum strategies and six weeks for contrarian strategies. In this study, most criteria, except for maximum drawdown, are ranked in the ascending order of the scoring criteria. After the assets are sorted, several ranking baskets are formed. In the S\&P 500 universe and the KOSPI 200 universe, we have ten ranking groups. Meanwhile, three ranking baskets are created for the SPDR U.S. sector ETFs because the number of assets in the ETF universe is much smaller than those of the previous two universes. The group one is for losers with the worst ranking scores, and the highest ranking group is for the best performers in the given selection rule. In other words, with the alternative selection rules based on maximum drawdown and recovery, loser groups gather much riskier assets and winner baskets consist of less riskier assets. This is the reason why maximum drawdown adopts the descending order but the others use the ascending order. Each ranking group is constructed as an equally-weighted portfolio with the assets in the group. For a monthly momentum (weekly contrarian) strategy, the winner group is on a long (short) position and the loser group is on a short (long) position. The size of the long position is exactly same with that of the short position in order to make the entire long-short portfolio be dollar-neutral. After the holding period of six months (weeks), each basket is liquidated. One-sixth of the portfolio is regularly constructed at the beginning of every month (week), i.e., it is an overlapping portfolio.

\subsection{Risk model for portfolio reward-risk profiles}
	Reward-risk measures for momentum/contrarian portfolio performance are calculated from daily time series of the overlapping portfolios. The maximum drawdown is directly computed from the empirical time series. The risk model for daily VaR, CVaR, and Sharpe ratio is the ARMA(1,1)-GARCH(1,1) model with classical tempered stable (CTS) innovations. This model is suggested by \cite{Kim:2010}, \cite{Kim:2010b}, and \cite{Kim:2011}, and several applications of the model are found in finance (\cite{Tsuchida:2012,Beck:2013}). The same model is also exploited for momentum portfolio construction (\cite{Choi:2015}) and diversified reward-risk parity portfolio construction (\cite{Choi:2021}). 
	
	The characteristic function of the CTS distribution is given by
	\begin{align}
		\phi(u)=&\exp\Big(ium-iu\Gamma(1-\alpha)(C_+\lambda_+^{1-\alpha}-C_-\lambda_-^{1-\alpha})\nonumber\\
		&+\Gamma(-\alpha)\big(C_+\big((\lambda_+-iu)^\alpha-\lambda_+^\alpha\big)+C_-\big((\lambda_-+iu)^\alpha-\lambda_-^\alpha\big)\big)
 \Big)\nonumber
	\end{align}
	where $m$ is the location parameter, $C_{\pm}$ are the scale parameters, $\alpha$ is the tail index, and $\lambda_{\pm}$ is the decay rates of upside/downside tails. All the CTS parameters are positive real numbers, in particular, $\alpha\in(0,2)$. From the viewpoint of risk management, $\alpha$ and $\lambda_{-}$ are important parameters. Smaller $\alpha$ values indicate heavier tails in the CTS distributions. Similarly, thicker downside tails are controlled by smaller $\lambda_-$.
	
	The procedures for calculating VaR, CVaR, and Sharpe ratio are the following steps (\cite{Kim:2011}). First of all, parameters of the ARMA-GARCH model with Student's $t$-innovations are estimated from maximum likelihood estimation (MLE). According to \cite{Rachev:2011}, residuals in the ARMA-GARCH model are considered as being generated from the probability distribution function of the CTS distribution which is obtained from the fast-Fourier transformation on the characteristic function of the CTS distribution. After then the parameter estimation of the CTS distribution is done by MLE. For detailed information on the CTS distribution and MLE, consult with \cite{Kim:2010}, \cite{Kim:2010b}, \cite{Kim:2011}, \cite{Rachev:2011}, and references therein.
	
\section{Performance and risk profiles of alternative strategies}
\label{sec_mddr_momentum_result}
\subsection{South Korea equity market: KOSPI 200}
\subsubsection{Weekly contrarian strategies}
	According to Table \ref{tbl_daily_summary_stat_risk_mddr_momentum_weekly_6_6_contrarian_kr_kp200}, ranking measures based on short-term recovery are robust prediction factors for cross-sectional contrarian signals in weekly scales in KOSPI 200 component universe. The recovery-based contrarian strategies are not only more profitable in average return but also less volatile in standard deviation than the traditional mean-reversion strategy achieving weekly 0.073\% on average under the volatility of 2.842\%. The best performer is the recovery portfolio with weekly 0.146\%, two-times larger than the average return of the original contrarian strategy. Additionally, stronger consistency in portfolio performance is guaranteed by much smaller standard deviation of 1.757\%, almost 40\%-decreased with respect to the volatility of the benchmark strategy. The CR strategy and the CMR strategy also obtain larger weekly average returns of 0.086\% and 0.078\% with reduced standard deviations of 2.665\% and 2.865\%, respectively. Moreover, kurtosis levels of these portfolios indicate that the portfolios are less exposed to extreme events. Meanwhile, maximum drawdown-related ranking rules such as the M, the CM, and the RM selection rules poorly predict directions of asset prices comparing with the benchmark measure.
	
\begin{sidewaystable}
\begin{center}
\caption{Summary statistics and risk measures of weekly 6/6 contrarian portfolios in South Korea KOSPI 200}
\resizebox{\textwidth}{!}{
\begin{tabular}{l l r r r r r r r r r}
\hline
Criterion & Portfolio & \multicolumn{5}{l}{Summary statistics} & \multicolumn{4}{l}{Risk measures}\\ \cline{3-7} \cline{8-11} 
 & & Mean & Std. Dev. & Skewness & Kurtosis & Fin. Wealth & Sharpe & $\textrm{VaR}_{95\%}$ & $\textrm{CVaR}_{95\%}$ & MDD \\ 
\hline
C& Winner (W) &0.1961&3.8032&-1.2946&6.7809&1.0119&0.0685&1.7519&2.5310&67.27\\
 & Loser (L) &0.2692&4.3576&-1.1792&7.6903&1.3889&0.0813&1.6808&2.5153&63.00\\
 & L - W &0.0731&2.8417&0.1864&1.8644&0.3770&0.0148&1.6864&2.4245&33.66\\
\\[-2ex] 
M& Winner (W) &0.2583&2.5125&-1.4858&7.4210&1.3330&0.0842&1.1302&1.6413&50.11\\
 & Loser (L) &0.2574&4.7328&-1.1544&7.0878&1.3281&0.0681&1.9362&2.7975&67.43\\
 & L - W &-0.0010&3.1694&-0.2546&2.7173&-0.0049&0.0140&1.7974&2.5476&60.08\\
\\[-2ex] 
R& Winner (W) &0.1688&4.0728&-1.5786&9.1786&0.8712&0.0603&1.6462&2.2375&72.17\\
 & Loser (L) &0.3143&3.3376&-1.4649&9.7503&1.6220&0.1003&1.2523&1.8700&51.78\\
 & L - W &0.1455&1.7567&0.0996&1.8461&0.7508&0.0465&1.1494&1.3907&30.09\\
\\[-2ex] 
RM& Winner (W) &0.2279&3.4532&-1.5951&8.4947&1.1762&0.0753&1.5765&2.1820&65.22\\
 & Loser (L) &0.2575&4.5200&-1.0493&6.8847&1.3286&0.0730&1.7822&2.6200&63.92\\
 & L - W &0.0295&2.7489&0.3269&2.2130&0.1524&0.0131&1.6210&2.3542&36.66\\
\\[-2ex] 
CM& Winner (W) &0.2258&3.4946&-1.2486&5.8310&1.1654&0.0737&1.6513&2.4019&61.34\\
 & Loser (L) &0.2594&4.5592&-1.1455&7.4612&1.3384&0.0780&1.8532&2.7334&64.35\\
 & L - W &0.0335&3.0442&0.1036&2.4027&0.1730&0.0165&1.8081&2.6182&41.05\\
\\[-2ex] 
CR& Winner (W) &0.1551&3.9277&-1.4040&7.7612&0.8003&0.0611&1.7428&2.4508&70.98\\
 & Loser (L) &0.2408&4.2109&-1.1877&8.4200&1.2424&0.0808&1.6678&2.4696&63.16\\
 & L - W &0.0857&2.6649&0.3887&1.4694&0.4421&0.0163&1.6105&2.2866&33.30\\
\\[-2ex] 
CMR& Winner (W) &0.1829&3.6266&-1.3528&6.8595&0.9439&0.0662&1.6819&2.3703&66.13\\
 & Loser (L) &0.2609&4.4422&-1.1052&7.4757&1.3461&0.0778&1.7475&2.5888&63.42\\
 & L - W &0.0779&2.8645&0.2827&2.0772&0.4022&0.0165&1.7426&2.5012&33.34\\
\hline
\end{tabular}
}\caption*{Summary statistics for the 6/6 weekly contrarian portfolios in South Korea KOSPI 200 are given in the table. Monthly average return, standard deviation, skewness, kurtosis and final wealth are found in the table. Additionally, risk measures for the 6/6 weekly contrarian portfolios in South Korea KOSPI 200 are also given. The risk measures are found from the daily performances. Sharpe ratio, VaR and CVaR are represented in daily percentage scale. Maximum drawdown (MDD) is in percentage scale.}

\label{tbl_daily_summary_stat_risk_mddr_momentum_weekly_6_6_contrarian_kr_kp200}
\end{center}
\end{sidewaystable}

	By adopting the R, the CR, and the CMR criteria, persistent profitability is also found at the levels of each ranking basket. First of all, loser groups in all the alternative contrarian portfolios perform as well as the loser group in cumulative return. Average returns of the long positions are in the range of 0.241\%--0.314\%, and return fluctuations of the long positions are less volatile than or comparable with that of the long basket in the benchmark portfolio. In particular, the loser group in the recovery measure achieves not only the largest profit but also the lowest deviation measure among all the loser baskets. Contrary to the recovery loser baskets, winner baskets in the R, the CR, and the CMR strategies yield less lucrative performance than the short baskets in the other strategies including the benchmark contrarian strategy. The poorer returns from the winner baskets in the contrarian long-short portfolios are also beneficial to the profitability of the entire portfolio.
	
	In Table \ref{tbl_daily_summary_stat_risk_mddr_momentum_weekly_6_6_contrarian_kr_kp200}, it is found that the outperformance in the recovery-related portfolios such as the R portfolio and the CR portfolio is achieved by taking less risks. These recovery strategies are less riskier in every risk measure than the strategies constructed from other ranking rules including the cumulative return. The R portfolio exhibits the largest Sharpe ratio, and the CR strategy is also one of the top portfolios in Sharpe ratio. Additionally, the lowest levels in VaR and CVaR are obtained by the R portfolio and the CR portfolio. In particular, the R strategy is exposed to the daily VaR of 1.149\% and the daily CVaR of 1.391\%, the smallest value in each risk measure. It is noticeable that the extent to which the CVaR value is decreased is more significant than that of the VaR measure. This fact indicates the existence of the much thinner downside tail in the performance of the R portfolio. Maximum drawdowns of the R portfolio and the CR portfolio are also lower than those of all the other strategies. Meanwhile, the maximum drawdown-related portfolios are exposed to worse risks with larger VaR and CVaR values.
	
	Each ranking basket in the R portfolio and the CR portfolio is also less riskier than the other competitive ranking groups as well as the benchmark. The winner groups and the loser groups of these two portfolios exhibit smaller VaRs and CVaRs than the corresponding ranking baskets in the benchmark. For example, the loser basket in the recovery criterion achieves the lowest VaR and CVaR values with 1.252\% and 1.870\%, respectively. The loser groups in the R strategy and the CR strategy obtain the top 2 largest Sharpe ratios among the other alternative strategies and the benchmark strategy. The maximum drawdown of the loser group in the recovery rule is also smaller than that of the long position in the traditional contrarian portfolio. For winner groups, the tendency is slightly weaker. Although short baskets are less riskier in VaR and CVaR, Sharpe ratios and maximum drawdowns of the short positions are worse than those of the winner basket in cumulative return. It is noteworthy that riskier short positions are more attractive for the profitability of the entire long-short portfolios. Opposite to the recovery-related strategies, the maximum drawdown strategies tend to construct less riskier short positions and much riskier long positions. These characteristics are not consistent with the desirable properties of contrarian portfolios.
	
\subsubsection{Monthly momentum strategies}
	In Table \ref{tbl_daily_summary_stat_risk_mddr_momentum_monthly_6_6_momentum_kr_kp200}, it is found that alternative momentum portfolios based on maximum drawdown-related stock selection rules outperform the traditional momentum portfolio  in a monthly scale. The best strategy is the momentum portfolio constructed from the composite ranking criterion with cumulative return and maximum drawdown. While the cumulative return criterion provides the trend-following strategy with the monthly return of 1.331\% and the standard deviation of 6.826\%, the CM portfolio achieves monthly 1.433\% on average with the volatility of 7.036\%. Additionally, the kurtosis of the CM portfolios is at the lowest level even comparing with the traditional momentum portfolio. The CMR portfolio and the RM portfolio are ranked at the next in performance by generating monthly returns of 1.311\% and 1.280\% under standard deviations of 6.729\% and 6.241\%, respectively. These portfolios are slightly worse in profitability but the volatility levels are also much more decreased with respect to that of the momentum strategy. Although the CR portfolio and the M portfolio underperform the benchmark, the strategies also exhibit steady returns. Contrary to the weekly strategies, the recovery criterion obtains the worst average return in a monthly scale.

\begin{sidewaystable}
\begin{center}
\caption{Summary statistics and risk measures of monthly 6/6 momentum portfolios in South Korea KOSPI 200}
\resizebox{\textwidth}{!}{
\begin{tabular}{l l r r r r r r r r r}
\hline
Criterion & Portfolio & \multicolumn{5}{l}{Summary statistics}& \multicolumn{4}{l}{Risk measures}\\ \cline{3-7} \cline{8-11} 
 & & Mean & Std. Dev. & Skewness & Kurtosis & Fin. Wealth & Sharpe & $\textrm{VaR}_{95\%}$ & $\textrm{CVaR}_{95\%}$ & MDD \\ 
\hline
C& Winner (W) &1.6292&8.7334&-0.4543&2.2045&1.8573&0.0775&1.7825&2.6373&64.48\\
 & Loser (L) &0.2987&8.9425&-0.8357&5.5529&0.3405&0.0641&1.9006&2.6312&65.46\\
 & W - L &1.3305&6.8258&0.0217&0.6760&1.5167&0.0545&2.5163&3.2597&59.87\\
\\[-2ex] 
M& Winner (W) &1.3075&5.5705&-0.5793&3.5011&1.4905&0.0885&1.1644&1.6563&46.13\\
 & Loser (L) &0.2841&9.5852&-0.8019&4.3932&0.3239&0.0591&2.0981&2.8616&67.36\\
 & W - L &1.0234&6.5769&-0.3515&1.0627&1.1667&0.0322&2.2287&2.7281&54.58\\
\\[-2ex] 
R& Winner (W) &1.3299&8.7707&-0.7226&3.1472&1.5161&0.0818&1.8927&2.8033&69.11\\
 & Loser (L) &0.9559&6.7652&-1.3312&6.0955&1.0898&0.0936&1.2107&1.7995&55.58\\
 & W - L &0.3740&4.0823&0.9101&2.2150&0.4264&0.0127&1.6553&2.3164&37.97\\
\\[-2ex] 
RM& Winner (W) &1.5416&7.5704&-0.2469&1.5551&1.7574&0.0828&1.7765&2.6339&58.39\\
 & Loser (L) &0.2613&9.3531&-0.7718&5.4615&0.2978&0.0613&2.0245&2.8005&67.25\\
 & W - L &1.2803&6.2406&-0.3327&1.5967&1.4596&0.0569&2.5490&3.2939&52.06\\
\\[-2ex] 
CM& Winner (W) &1.7005&8.0623&-0.3032&1.7154&1.9386&0.0785&1.6763&2.4660&59.37\\
 & Loser (L) &0.2676&9.1117&-0.8241&5.4601&0.3051&0.0635&1.9590&2.6871&66.38\\
 & W - L &1.4330&7.0357&-0.1969&0.5758&1.6336&0.0611&2.4930&3.1834&59.16\\
\\[-2ex] 
CR& Winner (W) &1.5449&8.8493&-0.5423&2.4918&1.7611&0.0778&1.8826&2.7816&67.88\\
 & Loser (L) &0.4421&8.6716&-0.8298&6.2348&0.5040&0.0643&1.7340&2.4558&63.74\\
 & W - L &1.1028&6.5956&0.2097&0.9879&1.2571&0.0404&2.3883&3.1240&62.37\\
\\[-2ex] 
CMR& Winner (W) &1.5557&8.2241&-0.3319&1.7862&1.7735&0.0754&1.8104&2.6763&62.70\\
 & Loser (L) &0.2451&9.1274&-0.8155&5.5803&0.2794&0.0625&1.9074&2.6450&66.39\\
 & W - L &1.3106&6.7290&-0.2114&0.7663&1.4941&0.0575&2.5282&3.2651&58.32\\
\hline
\end{tabular}
}\caption*{Summary statistics for the 6/6 monthly momentum portfolios in South Korea KOSPI 200 are given in the table. Monthly average return, standard deviation, skewness, kurtosis and final wealth are found in the table. Additionally, risk measures for the 6/6 monthly momentum portfolios in South Korea KOSPI 200 are also given. The risk measures are found from the daily performances. Sharpe ratio, VaR and CVaR are represented in daily percentage scale. Maximum drawdown (MDD) is in percentage scale.}
\label{tbl_daily_summary_stat_risk_mddr_momentum_monthly_6_6_momentum_kr_kp200}
\end{center}
\end{sidewaystable}

	Strong momentum in each ranking group basket of the momentum strategies associated with maximum drawdown is another evidence for robust prediction by the maximum drawdown criteria. Among all long baskets including the cumulative return winner, the strongest upside momentum of 1.701\% is achieved by the winner basket in the CM strategy. Additionally, the return volatility of 8.062\% for the CM winner basket is almost 10\%-smaller than that of the benchmark winner group. Similar to the CM strategy, the performance of the CMR winner basket is as lucrative as the benchmark winner basket, and the fluctuation of the performance is relatively decreased. Moreover, loser groups of all the maximum drawdown-based strategies underperform the traditional momentum loser group. The loser basket of the CM portfolio is less profitable in average return than the loser group in the cumulative return criterion. In addition, the CM loser is one of the worst baskets in average return among the loser groups. Comparing with the loser basket of the traditional momentum strategy, the RM strategy and the CMR strategy also exhibit stronger downside momentum in the short baskets which is advantageous for improving the profitability of the long-short portfolios. Meanwhile, the ranking baskets of the recovery-related portfolios do not exhibit the desirable characteristics of the momentum ranking groups.
	
	In Table \ref{tbl_daily_summary_stat_risk_mddr_momentum_monthly_6_6_momentum_kr_kp200}, risk profiles of the momentum portfolios indicate that the alternative portfolios ranked by the maximum drawdown-related selection rules are less riskier in various risk measures than the benchmark. For example, smaller VaR and CVaR levels are key properties of these alternative momentum strategies. Even for the RM portfolio and the CMR portfolio, the differences in risk measure are just in a few basis points. Additionally, smaller maximum drawdowns are characteristics of the alternative portfolios except for the CR portfolio. Moreover, higher Sharpe ratios are achieved by the CM, the CMR, and the RM strategies. From the observation, risk management can be improved by selecting the composite ranking criteria based on maximum drawdown.

	Each ranking group in the drawdown strategies also exhibits improved risk characteristics. Comparing with the long basket in the benchmark momentum portfolio, VaR, CVaR, and maximum drawdowns of the alternative winner baskets are reduced. Additionally, Sharpe ratios of these long baskets are increased with respect to that of the cumulative return winner group. The increased reward-risk measure and decreased risk metrics are desirable for the long positions. Meanwhile, worse VaRs, CVaRs, and maximum drawdowns of the loser groups indicate greater exposure to risks. Sharpe ratios of the alternative short baskets are weaker than that of the benchmark momentum loser. It is obvious that higher risk and poorer performance are more recommendable for short positions.

\subsection{U.S. equity market: SPDR sector ETFs}
\subsubsection{Weekly contrarian strategies}
	As shown in Table \ref{tbl_daily_summary_stat_risk_mddr_momentum_weekly_6_6_contrarian_us_sector_etf}, reversal found in the RM, the CR, and the CMR contrarian portfolios indicates that the short-term recovery is a robust forecasting measure for mean-reversion in the ETF universe. In particular, the CR selection rule is the best predictive factor on asset price among all the other contrarian strategies. The cumulative return-based portfolio is outperformed by the CR portfolio generating the weekly return of 0.094\%. Comparing with the benchmark, the less volatile performance of the CR strategy supports that the prediction based on the recovery-related selection rule is more consistent. Moreover, the CR portfolio is less exposed to extreme events than the benchmark. In addition to the CR strategy, the average return of the R strategy is not the best outcome but its standard deviation is at the lowest level, weekly 1.217\%. Although the other recovery-based ranking rules also exhibit good performance, the weekly returns are more volatile than the benchmark portfolio. 
	
\begin{sidewaystable}
\begin{center}
\caption{Summary statistics and risk measures of weekly 6/6 contrarian portfolios in US sector ETF}
\resizebox{\textwidth}{!}{
\begin{tabular}{l l r r r r r r r r r}
\hline
Criterion & Portfolio & \multicolumn{5}{l}{Summary statistics}  & \multicolumn{4}{l}{Risk measures}\\ \cline{3-7} \cline{8-11} 
 & & Mean & Std. Dev. & Skewness & Kurtosis & Fin. Wealth & Sharpe & $\textrm{VaR}_{95\%}$ & $\textrm{CVaR}_{95\%}$ & MDD \\ 
\hline
C& Winner (W) &0.0397&2.5424&-1.0070&5.6347&0.2876&0.0477&1.1783&1.4955&53.08\\
 & Loser (L) &0.1274&2.9360&-0.4862&7.3610&0.9226&0.0408&1.2427&1.5333&60.12\\
 & L - W &0.0877&1.7279&0.6594&7.4927&0.6350&0.0001&0.4469&0.6055&29.74\\
\\[-2ex] 
M& Winner (W) &0.0615&2.1450&-1.2742&7.7993&0.4454&0.0529&1.0917&1.4038&44.84\\
 & Loser (L) &0.1116&3.3237&-0.3631&5.7091&0.8083&0.0441&1.3652&1.7236&63.96\\
 & L - W &0.0501&2.1195&0.3290&5.5900&0.3629&0.0210&0.5570&0.7526&49.03\\
\\[-2ex] 
R& Winner (W) &0.0654&2.8563&-0.8552&6.0406&0.4735&0.0460&1.1853&1.5648&61.57\\
 & Loser (L) &0.1133&2.5224&-0.6937&6.5736&0.8206&0.0537&1.2192&1.4816&46.81\\
 & L - W &0.0479&1.2174&0.5730&3.7388&0.3470&0.0000&0.3403&0.4426&17.91\\
\\[-2ex] 
RM& Winner (W) &0.0467&2.3858&-1.0909&6.1818&0.3384&0.0511&1.1193&1.4491&52.07\\
 & Loser (L) &0.1158&3.1071&-0.3805&6.5959&0.8382&0.0408&1.2903&1.5781&60.78\\
 & L - W &0.0690&1.7563&0.7227&9.7128&0.4998&0.0000&0.4724&0.6073&33.84\\
\\[-2ex] 
CM& Winner (W) &0.0537&2.3661&-1.0646&5.4095&0.3891&0.0506&1.1356&1.4503&48.73\\
 & Loser (L) &0.1197&3.1199&-0.4035&6.9313&0.8663&0.0399&1.2742&1.5995&63.77\\
 & L - W &0.0659&1.8692&0.6222&8.1724&0.4772&0.0068&0.4846&0.6459&42.43\\
\\[-2ex] 
CR& Winner (W) &0.0335&2.6260&-1.0225&5.6197&0.2428&0.0458&1.1733&1.5236&57.61\\
 & Loser (L) &0.1270&2.8685&-0.4025&7.3544&0.9198&0.0408&1.3345&1.6153&56.89\\
 & L - W &0.0935&1.6477&0.7112&6.8603&0.6770&0.0000&0.4504&0.5750&20.46\\
\\[-2ex] 
CMR& Winner (W) &0.0463&2.4735&-0.9579&5.0344&0.3353&0.0496&1.1432&1.4590&53.24\\
 & Loser (L) &0.1208&3.0348&-0.3624&6.7464&0.8744&0.0398&1.3061&1.5975&61.50\\
 & L - W &0.0745&1.7891&0.7205&8.0367&0.5392&0.0000&0.5092&0.6519&32.00\\
\hline
\end{tabular}
}\caption*{Summary statistics for the 6/6 weekly contrarian portfolios in SPDR U.S. sector ETFs are given in the table. Monthly average return, standard deviation, skewness, kurtosis and final wealth are found in the table. Additionally, risk measures for the 6/6 weekly contrarian portfolios in SPDR U.S. sector ETFs are also given. The risk measures are found from the daily performances. Sharpe ratio, VaR and CVaR are represented in daily percentage scale. Maximum drawdown (MDD) is in percentage scale.}
\label{tbl_daily_summary_stat_risk_mddr_momentum_weekly_6_6_contrarian_us_sector_etf}
\end{center}
\end{sidewaystable}

	The portfolio-level reversal found in the recovery-based strategy is driven by the reversal from each ranking basket. The recovery-based measures predict both future winners and losers well. First of all, the winner baskets in the alternative portfolios underperform the loser baskets. Second, the loser baskets in the recovery strategies exhibit similar magnitudes of the reversion. In particular, the loser basket in the CR portfolio obtains the strongest weekly turn-around of 0.127\%. Additionally, the volatility of the long basket in the CR portfolio is the second-lowest one among all the alternative loser groups. Its strong mean-reversion is not limited to the loser group. For example, the average weekly return of the CR winner basket is 0.034\%, the worst performance among all the short baskets including the benchmark case. Moreover, the standard deviation of the weekly performance is at one of the highest levels. The long-short combination of the CR winner and the CR loser groups with the opposite characteristics makes the entire portfolio more lucrative.
	
	In Table \ref{tbl_daily_summary_stat_risk_mddr_momentum_weekly_6_6_contrarian_us_sector_etf}, it is found that the outperformance of the CR strategy is achieved while taking less risk. The CR portfolio also exhibits VaR, CVaR, and maximum drawdown of 0.450\%, 0.575\%, and 20.46\%, respectively. These risk measures are the second lowest numbers in VaR, CVaR and maximum drawdown among the alternative portfolios. The CVaR and the maximum drawdown of the portfolio are lower than the benchmark, and the VaR value is comparable with the risk measure of the traditional contrarian portfolio. Additionally, the recovery portfolio is the best portfolio in risk management for every risk measure such that its VaR, CVaR, and maximum drawdown are 0.340\%, 0.443\%, and 17.91\%. Comparing with the risk metrics of the traditional mean-reversion strategy, these measures are substantially reduced. Meanwhile, it is noteworthy that worse risk profiles are yielded by the other portfolios constructed from the maximum drawdown-related ranking rules.
	
	Similar to the entire long-short portfolio level, risk management in each ranking basket can be improved by considering the recovery measure in the stock selection process. In particular, the best loser group in performance and risk is constructed from the recovery criterion. All risk measures of the recovery loser basket are lower than those of any other loser groups in the alternative portfolios including the benchmark strategy. Moreover, the largest Shape ratio is also achieved by the loser in the recovery criterion. Opposite to the long basket, the winner group of the recovery portfolio exhibits the opposite characteristics, i.e., it obtains the worst risk metrics and Sharpe ratio. Since winner baskets in contrarian strategies are actually going short, riskier short positions are in general helpful to gain more profits for overall long-short portfolios.
	
\subsubsection{Monthly momentum strategies}
	Table \ref{tbl_daily_summary_stat_risk_mddr_momentum_monthly_6_6_momentum_us_sector_etf} reports that alternative stock selection rules constructed from maximum drawdown produce more improved forecasting on cross-sectional momentum phenomena in a monthly scale than the cumulative return criterion in the SPDR U.S. sector ETF universe. In particular, the maximum drawdown-based momentum portfolios outperform the cumulative momentum strategy. While the traditional momentum strategy generates monthly 0.117\% on average under the volatility of 3.552\%, the CMR strategy, the best portfolio among all criteria, achieves the monthly return of 0.172\% and the standard deviation of 3.565\%, i.e., the average performance is almost 50\%-increased and the standard deviation is increased less than 1\%. In the performance measure, the RM strategy and the CM strategy with monthly returns of 0.138\% and 0.121\%, respectively, are also followed by the benchmark strategy. These portfolios are also less volatile than the benchmark strategy. Contrary to the weekly strategies, the recovery-base strategies such as R strategy and the CR strategy are not more profitable than the cumulative return strategy. However, the monthly performances of the recovery portfolios are much less volatile.
	
\begin{sidewaystable}
\begin{center}
\caption{Summary statistics and risk measures of monthly 6/6 momentum portfolios in US sector ETF}
\resizebox{\textwidth}{!}{
\begin{tabular}{l l r r r r r r r r r}
\hline
Criterion & Portfolio & \multicolumn{5}{l}{Summary statistics}  & \multicolumn{4}{l}{Risk measures}\\ \cline{3-7} \cline{8-11} 
 & & Mean & Std. Dev. & Skewness & Kurtosis & Fin. Wealth & Sharpe & $\textrm{VaR}_{95\%}$ & $\textrm{CVaR}_{95\%}$ & MDD \\ 
\hline
C& Winner (W) &0.3742&4.5444&-0.7738&1.3970&0.6062&0.0518&1.3639&1.7161&48.23\\
 & Loser (L) &0.2568&5.3434&-0.6718&1.6714&0.4160&0.0373&1.4811&1.7675&65.48\\
 & W - L &0.1174&3.5522&-0.1950&1.6853&0.1903&0.0001&0.5529&0.7246&39.55\\
\\[-2ex] 
M& Winner (W) &0.2820&3.6813&-0.8993&2.3732&0.4569&0.0540&1.2180&1.5566&44.57\\
 & Loser (L) &0.1869&5.7772&-0.4749&0.9533&0.3027&0.0361&1.6286&1.9842&64.60\\
 & W - L &0.0951&3.5423&-0.1730&1.7801&0.1541&-0.0025&0.7813&1.0325&38.14\\
\\[-2ex] 
R& Winner (W) &0.3616&5.0347&-0.7006&1.7317&0.5859&0.0491&1.5282&1.9039&59.12\\
 & Loser (L) &0.2463&4.3456&-0.6546&1.3965&0.3990&0.0452&1.3462&1.6252&49.28\\
 & W - L &0.1154&2.7612&-0.2002&0.9125&0.1869&0.0001&0.3750&0.5204&29.21\\
\\[-2ex] 
RM& Winner (W) &0.3568&4.0697&-0.9282&1.9866&0.5780&0.0488&1.2738&1.6263&46.52\\
 & Loser (L) &0.2191&5.4810&-0.5295&1.1879&0.3549&0.0371&1.4805&1.8016&63.39\\
 & W - L &0.1377&3.2694&-0.4041&3.0940&0.2231&-0.0000&0.5050&0.6658&35.87\\
\\[-2ex] 
CM& Winner (W) &0.3810&4.2173&-0.8257&1.6097&0.6172&0.0458&1.2515&1.6113&44.80\\
 & Loser (L) &0.2599&5.4694&-0.5744&1.2093&0.4210&0.0322&1.5393&1.8406&63.52\\
 & W - L &0.1211&3.4904&-0.3021&2.1704&0.1962&0.0000&0.6991&0.8904&37.85\\
\\[-2ex] 
CR& Winner (W) &0.4015&4.6474&-0.7771&1.4298&0.6504&0.0541&1.4251&1.8317&50.05\\
 & Loser (L) &0.2911&5.0446&-0.5675&1.6271&0.4716&0.0420&1.3626&1.6324&61.33\\
 & W - L &0.1104&3.2966&-0.3510&1.2258&0.1788&0.0001&0.3790&0.4868&36.23\\
\\[-2ex] 
CMR& Winner (W) &0.4009&4.3287&-0.8553&1.5424&0.6495&0.0469&1.2709&1.6388&46.38\\
 & Loser (L) &0.2286&5.4417&-0.5751&1.2593&0.3703&0.0331&1.5221&1.8123&64.15\\
 & W - L &0.1724&3.5652&-0.3392&1.9933&0.2792&0.0000&0.6383&0.7834&40.65\\
\hline
\end{tabular}
}\caption*{Summary statistics for the 6/6 monthly momentum portfolios in SPDR U.S. sector ETFs are given in the table. Monthly average return, standard deviation, skewness, kurtosis and final wealth are found in the table. Additionally, risk measures for the 6/6 monthly momentum portfolios in SPDR U.S. sector ETFs are also given. The risk measures are found from the daily performances. Sharpe ratio, VaR and CVaR are represented in daily percentage scale. Maximum drawdown (MDD) is in percentage scale.}

\label{tbl_daily_summary_stat_risk_mddr_momentum_monthly_6_6_momentum_us_sector_etf}
\end{center}
\end{sidewaystable}

	Regardless of criterion, predictability on cross-sectional momentum is still persistent in each ranking basket, i.e., winners tend to outperform losers in a monthly scale. Winner groups of the alternative portfolios are as consistently lucrative in average return as the winner group of the traditional momentum strategy. In particular, the strongest momentum at the ranking basket levels is achieved by the CMR strategy. The long (short) basket of the CMR portfolio outperforms (underperforms) the corresponding basket of the benchmark portfolio. Strong trend-following anomaly can also be found in winner and loser groups of the CM portfolio and the RM portfolio. The same pattern is observed in the cases of the other recovery-related ranking portfolios. For example, the R criterion and the CR criterion also provide meaningful momentum at the ranking group levels.
	
	In Table \ref{tbl_daily_summary_stat_risk_mddr_momentum_monthly_6_6_momentum_us_sector_etf}, the alternative strategies are less riskier than the traditional momentum strategy. First of all, maximum drawdowns for the portfolios are lower than or comparable with the benchmark case. Additionally, lower VaR and CVaR levels found in the alternative portfolios indicate that the recovery-related selection rules are superior in avoiding severe losses in portfolio performance. In particular, the RM portfolio exhibits lower risk measures while achieving the better monthly performance. The reward-risk ratio of the CMR portfolio is also improved due to the almost 50\%-increased monthly return and the slightly increased risk measure. The R portfolio and the CR portfolio are not only substantially less riskier but also as profitable as the trend-following strategy by cumulative return. 

	Risk measures of ranking baskets in the alternative maximum drawdown momentum strategies are improved. Comparing with the benchmark cases, decreased (increased) VaR and CVaR levels for winner (loser) groups are observed in the cases of the maximum drawdown portfolios such as the M, the RM, the CM, and the CMR portfolios. Similarly, smaller (larger) maximum drawdowns of the winner (loser) groups are obtained. Meanwhile, the opposite pattern is found at the ranking baskets in the R portfolio and CR portfolio. The strategies yield larger (smaller) VaR and CVaR values for winner (loser) groups than the cumulative return strategy. Long positions in the recovery-related momentum strategies are exposed to more severe risks and it is not advisable for the characteristics of long positions.

\subsection{U.S. equity market: S\&P 500}
\subsubsection{Weekly contrarian strategies}
	Similar to the weekly contrarian strategies in the universes covered above, recovery measures provide more robust prediction on weekly mean-reversion in S\&P 500 universe. In Table \ref{tbl_daily_summary_stat_risk_mddr_momentum_weekly_6_6_contrarian_us_spx}, it is found that the R portfolio and the CR portfolio strongly outperform the traditional contrarian portfolio. In particular, the recovery criterion constructs the best portfolio in both performance and volatility, achieving the weekly return of 0.087\% and the weekly standard deviation of 1.785\%. It is noteworthy that these statistics are significantly improved with respect to the benchmark case. The CR strategy is also followed by the cumulative return-based mean-reversion strategy, and its second-smallest volatility imposes that the forecasting on reversal is consistent. Although the CMR portfolio is also attractive in profitability, the performance of the portfolio, placed behind that of the C portfolio, is more volatile than the R portfolio and the CR portfolio. Meanwhile, maximum drawdown-based strategies deliver poorer returns than the benchmark.
	
\begin{sidewaystable}
\begin{center}
\caption{Summary statistics and risk measures of weekly 6/6 contrarian portfolios in US S\&P 500}
\resizebox{\textwidth}{!}{
\begin{tabular}{l l r r r r r r r r r}
\hline
Criterion & Portfolio & \multicolumn{5}{l}{Summary statistics}  & \multicolumn{4}{l}{Risk measures}\\ \cline{3-7} \cline{8-11} 
 & & Mean & Std. Dev. & Skewness & Kurtosis & Fin. Wealth & Sharpe & $\textrm{VaR}_{95\%}$ & $\textrm{CVaR}_{95\%}$ & MDD \\ 
\hline
C& Winner (W) &0.1113&3.0677&-0.9314&6.8064&1.1541&0.0446&1.4100&1.7737&63.77\\
 & Loser (L) &0.1533&4.1437&-0.0867&10.6371&1.5894&0.0241&1.3656&1.6937&81.06\\
 & L - W &0.0420&2.8100&0.7533&11.9326&0.4353&0.0033&0.5483&0.7547&72.56\\
\\[-2ex] 
M& Winner (W) &0.0746&1.7631&-1.3442&9.6682&0.7732&0.0581&0.9729&1.2641&42.75\\
 & Loser (L) &0.0681&4.8649&-0.0346&7.9994&0.7059&0.0155&1.5204&1.9079&87.78\\
 & L - W &-0.0065&4.0948&0.3949&9.1759&-0.0673&0.0002&0.8300&1.1168&95.23\\
\\[-2ex] 
R& Winner (W) &0.1018&3.6468&-0.5561&7.9410&1.0554&0.0227&1.4296&1.8275&72.18\\
 & Loser (L) &0.1890&2.6504&-0.5853&9.1379&1.9595&0.0600&1.2035&1.5569&60.05\\
 & L - W &0.0872&1.7852&0.1912&12.1168&0.9041&0.0110&0.4457&0.5891&43.99\\
\\[-2ex] 
RM& Winner (W) &0.1006&2.5486&-1.0325&6.7681&1.0428&0.0450&1.3311&1.6871&57.66\\
 & Loser (L) &0.1124&4.4562&-0.1119&9.3392&1.1653&0.0252&1.4218&1.7835&84.29\\
 & L - W &0.0118&3.0944&0.4860&14.7853&0.1225&0.0001&0.5305&0.7457&84.48\\
\\[-2ex] 
CM& Winner (W) &0.1040&2.5745&-1.0275&5.5904&1.0785&0.0510&1.3392&1.6716&57.59\\
 & Loser (L) &0.1082&4.4531&-0.1169&9.5164&1.1219&0.0196&1.4254&1.7867&84.00\\
 & L - W &0.0042&3.2824&0.5567&13.2651&0.0433&0.0029&0.5902&0.8245&87.54\\
\\[-2ex] 
CR& Winner (W) &0.1079&3.2285&-0.7571&6.7366&1.1188&0.0372&1.4321&1.8091&65.75\\
 & Loser (L) &0.1848&3.9246&0.0033&11.0980&1.9166&0.0274&1.3695&1.6917&78.68\\
 & L - W &0.0769&2.5635&0.7710&10.9561&0.7979&0.0070&0.5445&0.7524&60.46\\
\\[-2ex] 
CMR& Winner (W) &0.1017&2.8001&-1.0088&6.4674&1.0549&0.0442&1.3960&1.7511&60.12\\
 & Loser (L) &0.1309&4.3011&-0.0817&10.2070&1.3571&0.0210&1.4003&1.7510&82.92\\
 & L - W &0.0291&3.0098&0.6234&14.0672&0.3023&0.0057&0.5378&0.7453&80.23\\
\hline
\end{tabular}
}\caption*{Summary statistics for the 6/6 weekly momentum portfolios in S\&P 500 are given in the table. Monthly average return, standard deviation, skewness, kurtosis and final wealth are found in the table. Additionally, risk measures for the 6/6 weekly contrarian portfolios in U.S. S\&P 500 are also given. The risk measures are found from the daily performances. Sharpe ratio, VaR and CVaR are represented in daily percentage scale. Maximum drawdown (MDD) is in percentage scale.}
\label{tbl_daily_summary_stat_risk_mddr_momentum_weekly_6_6_contrarian_us_spx}
\end{center}
\end{sidewaystable}

	Mean-reversion is also observed at ranking baskets of the alternative recovery rules. In particular, the losers in recovery outperform the winners in the same measure and the loser group in the benchmark measure. First of all, the loser basket of the R portfolio achieves not only the highest weekly return of 0.189\% but also the smallest volatility of 2.650\% among all the loser groups including the cumulative return loser group. In addition to the recovery criterion, the CR loser group also exhibits 0.185\% on average, the second-largest average return, with the second-smallest standard deviation. Opposite to the long baskets, the alternative winner baskets gain poorer profits than the benchmark winner basket, and the worst performance is obtained by the winner group in maximum drawdown. In addition, the winner baskets in the R criterion and the CR criterion suffer from relatively larger fluctuations in return.

	According to the risk measures in Table \ref{tbl_daily_summary_stat_risk_mddr_momentum_weekly_6_6_contrarian_us_spx}, it is possible to construct low-risk contrarian portfolios based on the recovery-based measures. The alternative recovery portfolios are more robust for avoiding severe risks than the traditional reversal portfolio and the maximum drawdown portfolios. All the recovery strategies are less riskier in VaR and CVaR. Additionally, Sharpe ratios are improved by the recovery-based portfolios such as the R, the CMR, and the CR portfolios. Moreover, lower maximum drawdowns than that of the cumulative return strategy are obtained only by the R portfolio and the CR portfolio. In particular, the maximum drawdown of the R portfolio is at the lowest level, 40\%-decreased with respect to the benchmark. Meanwhile, the portfolios constructed from the selection rules stemming from maximum drawdown are exposed to higher risks. For example, the reward-risk measures of the M strategy are ranked at the worst level.
	
	In each ranking basket, the ranking baskets by the recovery-related measures are also more compatible with desirable risk profiles for contrarian strategies. For example, the loser groups of the R portfolio and the CR portfolio achieve the lowest VaRs, CVaRs, and maximum drawdowns. Additionally, the Sharpe ratio of the long basket in the recovery portfolio is significantly larger than the other cases including the benchmark loser group. In addition, the second largest Sharpe ratio among the loser groups is produced by the CR criterion. Meanwhile, risk exposures of the R winner group and the CR winner group are most significant. The VaR, CVaR, and maximum drawdown of the winner basket in the recovery portfolio are worse than those of the benchmark winner basket. The short baskets of the R portfolio and the CR portfolio are also the weakest ranking groups in Sharpe ratio. Opposite to the ranking baskets of the recovery-based strategies, the long (short) baskets by the maximum drawdown-related measures exhibit larger (smaller) risks than the benchmark long (short) basket and the alternative short (long) baskets. This pattern contributes to the underperformance of the maximum drawdown-related strategies.
	
\subsubsection{Monthly momentum strategies}
	Contrary to the weekly strategies, maximum drawdown measures improve predictability on price momentum in the S\&P 500 universe. Table \ref{tbl_daily_summary_stat_risk_mddr_momentum_monthly_6_6_momentum_us_spx} shows that maximum drawdown-related momentum portfolios exhibit more profitable performance than other portfolios. In particular, the CM portfolio, the best performer among all the alternative strategies, achieves the monthly return of 0.498\% and the volatility of 7.303\%, while the monthly return of 0.480\% and the standard deviation of 6.835\% are obtained by the traditional momentum strategy. Moreover, the CM portfolio is less exposed to extreme events with the lower kurtosis than the benchmark. The CMR portfolio of monthly 0.483\% on average is also followed in the performance measure by the benchmark strategy. Opposite to the drawdown portfolios, poorer performances with smaller standard deviations are gained by the recovery-associated portfolios such as the R, the RM and the CR portfolios. In particular, momentum yielded by the recovery criterion is negligible.

\begin{sidewaystable}
\begin{center}
\caption{Summary statistics and risk measures of monthly 6/6 momentum portfolios in US S\&P 500}
\resizebox{\textwidth}{!}{
\begin{tabular}{l l r r r r r r r r r}
\hline
Criterion & Portfolio & \multicolumn{5}{l}{Summary statistics}  & \multicolumn{4}{l}{Risk measures}\\ \cline{3-7} \cline{8-11} 
 & & Mean & Std. Dev. & Skewness & Kurtosis & Fin. Wealth & Sharpe & $\textrm{VaR}_{95\%}$ & $\textrm{CVaR}_{95\%}$ & MDD \\ 
\hline
C& Winner (W) &0.8494&5.4621&-0.4642&1.4226&1.9877&0.0520&1.5696&1.8195&56.70\\
 & Loser (L) &0.3698&8.5188&0.0213&2.6105&0.8654&0.0212&1.7371&1.8518&80.28\\
 & W - L &0.4796&6.8349&-0.9646&6.1612&1.1223&0.0078&1.1517&1.5101&65.05\\
\\[-2ex] 
M& Winner (W) &0.5583&3.1605&-0.9713&2.0259&1.3063&0.0661&1.1149&1.4730&42.61\\
 & Loser (L) &0.2909&9.3503&-0.0700&1.6857&0.6806&0.0003&1.8322&1.9714&82.88\\
 & W - L &0.2674&8.0069&-0.5832&3.5287&0.6257&-0.0000&1.6163&1.8233&69.40\\
\\[-2ex] 
R& Winner (W) &0.7122&6.4972&-0.4567&1.4513&1.6666&0.0373&1.6199&1.8267&72.42\\
 & Loser (L) &0.6974&5.1412&-0.6791&2.4862&1.6320&0.0452&1.2706&1.4728&59.28\\
 & W - L &0.0148&3.7651&1.0177&8.9110&0.0346&0.0132&0.5408&0.6684&54.38\\
\\[-2ex] 
RM& Winner (W) &0.7306&4.4157&-0.7131&1.7078&1.7096&0.0572&1.4656&1.7365&52.14\\
 & Loser (L) &0.3366&9.0230&-0.0406&2.1542&0.7878&0.0003&1.7469&1.8615&82.67\\
 & W - L &0.3939&6.9933&-1.0825&6.0699&0.9218&0.0070&1.1785&1.4420&66.01\\
\\[-2ex] 
CM& Winner (W) &0.8100&4.7207&-0.5439&1.7188&1.8955&0.0596&1.5040&1.7783&53.69\\
 & Loser (L) &0.3120&8.9593&-0.0229&2.1164&0.7302&0.0132&1.7878&1.8978&82.82\\
 & W - L &0.4980&7.3026&-0.9736&5.3991&1.1653&0.0086&1.2913&1.6035&67.73\\
\\[-2ex] 
CR& Winner (W) &0.7926&5.7039&-0.4881&1.2752&1.8546&0.0502&1.6019&1.8395&60.49\\
 & Loser (L) &0.4182&8.0488&0.0221&2.9937&0.9785&0.0266&1.6870&1.8152&78.66\\
 & W - L &0.3744&6.2013&-0.9496&7.0288&0.8761&0.0002&0.9677&1.3117&60.28\\
\\[-2ex] 
CMR& Winner (W) &0.8277&5.0102&-0.5146&1.7918&1.9367&0.0575&1.5411&1.8006&55.17\\
 & Loser (L) &0.3450&8.7546&-0.0142&2.3770&0.8073&0.0176&1.7602&1.8713&81.68\\
 & W - L &0.4827&6.9875&-1.0387&6.4780&1.1295&0.0116&1.1821&1.5230&65.58\\
\hline
\end{tabular}
}\caption*{Summary statistics for the 6/6 monthly momentum portfolios in S\&P 500 are given in the table. Monthly average return, standard deviation, skewness, kurtosis and final wealth are found in the table. Additionally, risk measures for the 6/6 monthly momentum portfolios in U.S. S\&P 500 are also given. The risk measures are found from the daily performances. Sharpe ratio, VaR and CVaR are represented in daily percentage scale. Maximum drawdown (MDD) is in percentage scale.}
\label{tbl_daily_summary_stat_risk_mddr_momentum_monthly_6_6_momentum_us_spx}
\end{center}
\end{sidewaystable}

	Strong momentum by the alternative maximum drawdown measures is also observed in ranking baskets. In particular, the outperformance of the maximum drawdown momentum strategies is originated from the underperformance found in the short positions. Most loser baskets in the maximum drawdown portfolios are less profitable in average return than the traditional momentum short position. Except for the R strategy and the CR strategy, the short baskets of the composite ranking criteria gain monthly 0.291\%--0.345\%, while the cumulative return loser group obtains monthly 0.370\% on average. When the loser baskets are in short-selling, larger portions of the relative profits are generated by the underperformance in the short positions. Regardless of criterion, the less volatile returns of the alternative long baskets indicate that the prediction by the drawdown measures is more consistent. Although the profitability of the winner groups is slightly worse than the average return of 0.849\% from the traditional momentum winner group, the profits from the winners are in the comparable range of 0.712\%--0.823\% with the benchmark winner. The exception is the M-strategy which earns monthly 0.558\% from the long position. Additionally, the winner baskets in the maximum drawdown-related measures exhibit less volatile historical returns than the recovery or the cumulative return winner groups.
	
	The superiority of the maximum drawdown momentum strategies is supported by higher Sharpe ratios. According to reward-risk measures of the portfolios given in Table \ref{tbl_daily_summary_stat_risk_mddr_momentum_monthly_6_6_momentum_us_spx}, Sharpe ratios of the RM, the CM and the CMR portfolios are comparable with or greater than the Sharpe ratio of the cumulative return-based portfolio. Meanwhile, the tendency that outperformance is achieved by the alternative portfolios while taking low downside risks is not as strong as the portfolios implemented in other universes. The alternative portfolios tend to yield larger VaR and CVaR measures than the benchmark strategy. Additionally, maximum drawdown also exhibits the similar pattern with VaR and CVaR.
	
	Risk measures of each alternative ranking basket are consistent with the outperformance of the maximum drawdown momentum strategies. Lower (higher) VaR, CVaR, and maximum drawdown levels are achieved by winner (loser) groups in the maximum drawdown-related stock selection rules. Its long position is less riskier than that of the benchmark momentum strategy while the opposite position is under greater risks of losing money. Additionally, larger (smaller) Sharpe ratios are obtained by the winner (loser) groups of the strategies. These opposite behaviors make the entire portfolios be profitable. Meanwhile, the recovery-related portfolios such as the R portfolio and the CR portfolio exhibit higher (lower) VaR, CVaR, and maximum drawdown for the winner (loser) groups.
	
\subsection{Overall results}
	Regardless of asset class and market, alternative selection rules using maximum drawdown and consecutive recovery achieve more enhanced predictability on profitability than the traditional momentum and contrarian portfolios. In particular, exploiting composite ranking rules with cumulative return, maximum drawdown, and recovery measures is superior to adopting only a single criterion.
	
	\begin{figure}[h!]
		\subfigure[Weekly in South Korea KOSPI 200]{\includegraphics[width=6cm]{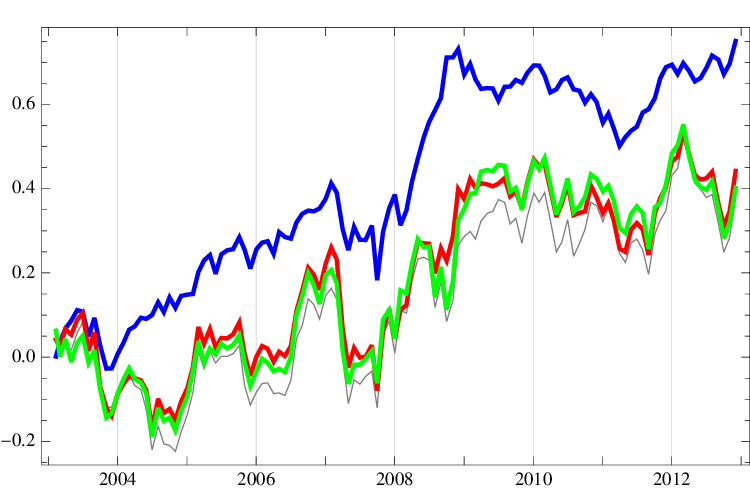}}
		\subfigure[Monthly in South Korea KOSPI 200]{\includegraphics[width=6cm]{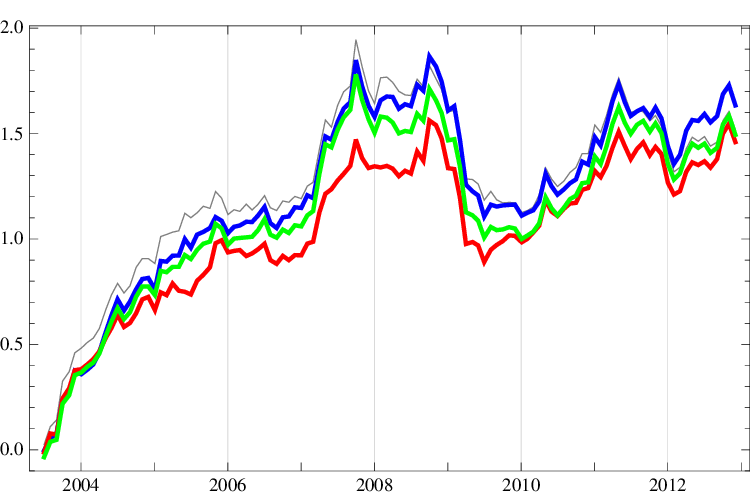}}
		\subfigure[Weekly in SPDR U.S. sector ETF]{\includegraphics[width=6cm]{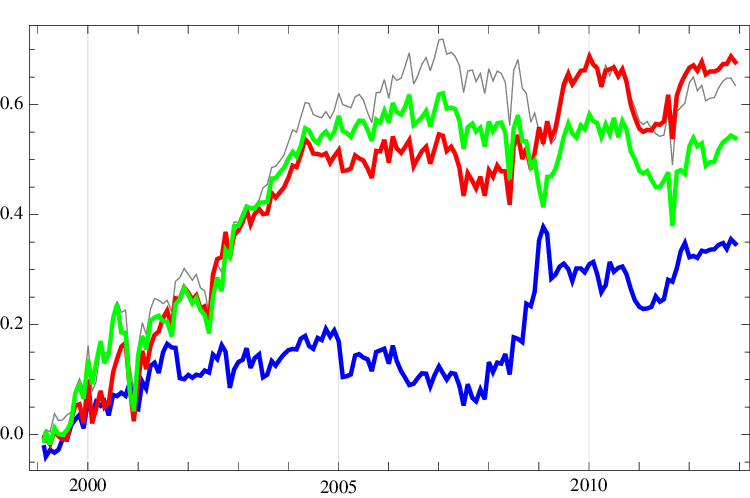}}
		\subfigure[Monthly in SPDR U.S. sector ETF]{\includegraphics[width=6cm]{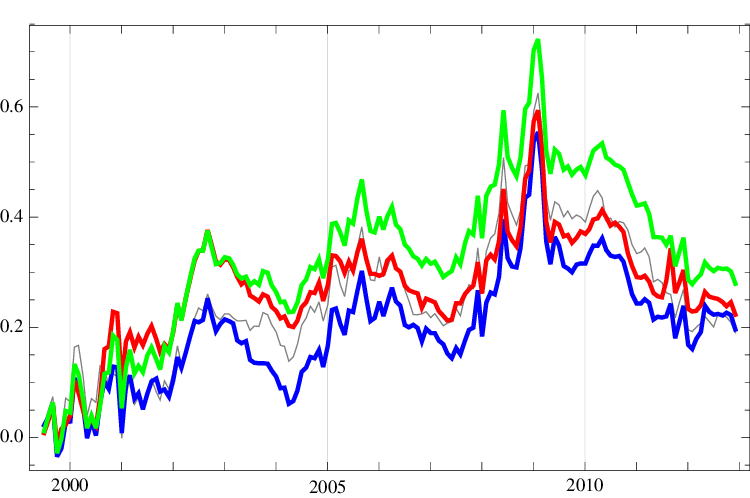}}
		\subfigure[Weekly in U.S. S\&P 500]{\includegraphics[width=6cm]{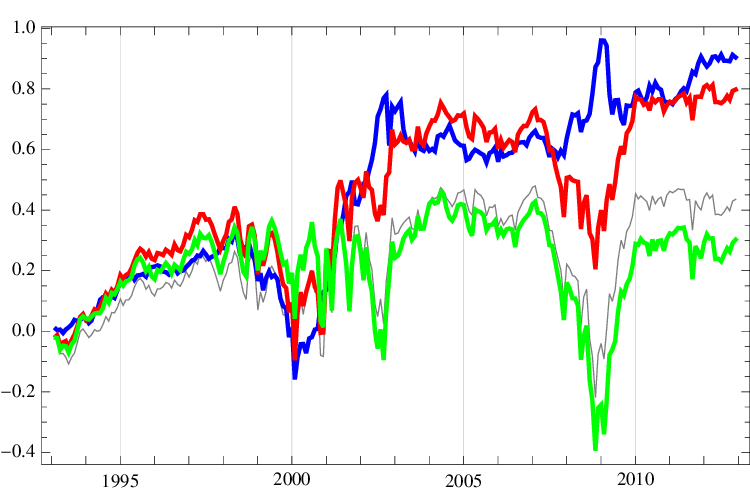}}
		\subfigure[Monthly in U.S. S\&P 500]{\includegraphics[width=6cm]{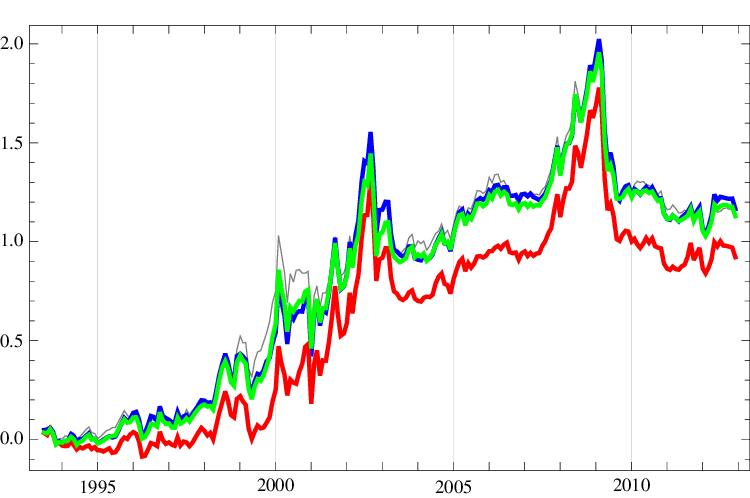}}
		\caption{For weekly contrarian, cumulative returns for the traditional contrarian (light gray), R (blue), CR (red) and CMR (green) portfolios. For monthly momentum, cumulative returns for the traditional momentum (light gray), CM (blue), RM (red) and CMR (green) portfolios.}
		\label{grp_acc_return_mddr_momentum_6_6}
\end{figure}

	In a weekly scale, the contrarian portfolios constructed by the recovery-based measures exhibit outperformance in profitability over the traditional contrarian portfolio. The R, the CR and the CMR criteria are the best stock selection rules for detecting weekly mean-reversion in many markets. High profitability and low volatility are achieved by the alternative portfolios. Historical cumulative returns of these portfolios are found in the left column of Figure \ref{grp_acc_return_mddr_momentum_6_6}.
	
	For monthly momentum portfolio construction, maximum drawdown-related measures are the best ranking criteria. The CM criterion provides the portfolio performing well across three different markets. Future winners and losers are also well-predicted by the CMR and the RM selection rules. Cumulative returns for these portfolios are also given in the right column of Figure \ref{grp_acc_return_mddr_momentum_6_6}.
	 
	It is noteworthy that the best criteria for the strategies are related to time scales and type of the strategy. For momentum, the maximum drawdown should be minimized for winners but maximized for losers because the maximum drawdown penalizes the long-term momentum trends. Meanwhile, the recovery during short-term mean-reversion should be decreased for losers. If the recovery during the past period is too large for the loser, it is considered that the asset already exhausted the fuel for the reversion, and it is hard to keep the turn-around trends.
	
	The excellence in profitability and volatility of the alternative portfolios is achieved with taking less risks. While outperforming the benchmark portfolio, the alternative portfolios are less riskier in VaR, CVaR, and maximum drawdown. The high-profit but low-risk aspect is consistent with the low-volatility anomaly (\cite{Blitz:2007, Baker:2011}). It is also interesting that the risk measures of each ranking group are also well-matched to the purpose of portfolio construction, i.e., winner (loser) groups in a monthly scale are less (much) riskier than the corresponding basket of the benchmark momentum. The ranking groups of the weekly contrarian strategies exhibit the opposite characteristics which are also consistent with the goal of the contrarian portfolio.

\section{Factor analysis}	
\label{sec_mddr_momentum_factor_analysis}

	As shown in the previous section, maximum drawdown and consecutive recovery are useful stock selection rules for acquiring more portfolio profits with taking less risks in various time scales and markets. 
	
	The factor analysis with the Carhart four factors (\cite{Carhart:1997}) is a robust way to test the outperformance of the alternative strategies. The Carhart factor model decomposes portfolio returns with respect to market factor (MKT), size factor (SMB), value factor (HML), and momentum factor (MOM). Among these factors, the market factor, the size factor, and the value factor are the three factors in \cite{Fama:1996}. The portfolio return $r_p$ is regressed with the Carhart four factors\footnote{The data for the four factors are downloaded from K. R. French's data library at Dartmouth.}:
	\begin{equation}
		r_p=\alpha+\beta_{MKT} f_{MKT}+\beta_{SMB} f_{SMB}+\beta_{HML} f_{HML}+\beta_{MOM} f_{MOM}+\epsilon_p\nonumber
	\end{equation}
	where $\epsilon_p$ is the residual, $\alpha$ is the intercept, and $\beta_i$ is the $i$-th factor in the Carhart model. For monthly momentum strategies, the daily performance of a portfolio is accumulated to monthly return $r_p$. For weekly contrarian strategies, the daily returns are converted to weekly returns. The factor returns are also given in the corresponding time scales. In this section, we focus on the alternative portfolios in the S\&P 500 universe.
	
\subsection{Weekly contrarian strategies}
	Regression intercepts and factor exposures of the alternative contrarian strategies are given in Table \ref{tbl_carhart_regression_mddr_momentum_weekly_6_6_contrarian_us_spx}. The factor analysis shows that the recovery measure is useful to detect meaningful short-term reversion. The R portfolio and the CR portfolio achieve not only positive Carhart four-factor alphas but also greater intercepts than that of the traditional contrarian portfolio. Moreover, the only statistically-significant alpha is obtained by the recovery strategy. 	
	
\begin{table}[h!]
\begin{center}
\caption{Carhart 4-factor regression of weekly 6/6 contrarian portfolios in US S\&P 500}
\resizebox{\textwidth}{!}{
\begin{tabular}{l l r @{} l r @{} l r @{} l r @{} l r @{} l r}
\hline
Criterion & Portfolio & \multicolumn{11}{l}{Factor loadings} \\ \cline{3-13} 
 & & \multicolumn{2}{c}{$\alpha(\%)$} & \multicolumn{2}{c}{$\beta_{MKT}$} & \multicolumn{2}{c}{$\beta_{SMB}$} & \multicolumn{2}{c}{$\beta_{HML}$} & \multicolumn{2}{c}{$\beta_{MOM}$} & $R^2$ \\ 
\hline
C&Winner (W)&-0.0629&&0.9853&${}^{**}$&0.1314&${}^{*}$&0.2960&${}^{**}$&0.1363&${}^{**}$&0.8768\\ 
&Loser (L)&-0.0700&&1.7110&${}^{**}$&0.4101&${}^{**}$&0.6486&${}^{**}$&-0.2663&${}^{**}$&0.8496\\ 
&L -- W&-0.0071&&0.7257&${}^{**}$&0.2787&${}^{*}$&0.3525&${}^{**}$&-0.4027&${}^{**}$&0.1934\\ 
\\[-2ex]
M&Winner (W)&-0.0018&&0.6092&${}^{**}$&-0.2030&${}^{**}$&-0.1196&${}^{**}$&0.0652&${}^{**}$&0.8491\\ 
&Loser (L)&-0.1570&&1.9296&${}^{**}$&0.6701&${}^{**}$&1.0841&${}^{**}$&-0.3291&${}^{**}$&0.8694\\ 
&L -- W&-0.1552&&1.3204&${}^{**}$&0.8731&${}^{**}$&1.2037&${}^{**}$&-0.3943&${}^{**}$&0.6927\\ 
\\[-2ex]
R&Winner (W)&-0.0974&&1.3647&${}^{**}$&0.2942&${}^{**}$&0.6562&${}^{**}$&-0.0209&&0.9081\\ 
&Loser (L)&0.0400&&1.1615&${}^{**}$&0.0903&${}^{**}$&0.1138&${}^{**}$&-0.0912&${}^{**}$&0.9473\\ 
&L -- W&0.1373&${}^{*}$&-0.2032&${}^{**}$&-0.2039&${}^{**}$&-0.5423&${}^{**}$&-0.0703&&0.4523\\ 
\\[-2ex]
RM&Winner (W)&-0.0310&&0.8872&${}^{**}$&0.0445&&0.1116&${}^{**}$&0.0878&${}^{*}$&0.8760\\ 
&Loser (L)&-0.1431&&1.8341&${}^{**}$&0.5219&${}^{**}$&0.7990&${}^{**}$&-0.2927&${}^{**}$&0.8697\\ 
&L -- W&-0.1122&&0.9469&${}^{**}$&0.4774&${}^{**}$&0.6874&${}^{**}$&-0.3805&${}^{**}$&0.4566\\ 
\\[-2ex]
CM&Winner (W)&-0.0426&&0.7996&${}^{**}$&0.0113&&0.0504&&0.1559&${}^{**}$&0.8567\\ 
&Loser (L)&-0.1218&&1.8218&${}^{**}$&0.5250&${}^{**}$&0.8167&${}^{**}$&-0.3006&${}^{**}$&0.8609\\ 
&L -- W&-0.0792&&1.0221&${}^{**}$&0.5137&${}^{**}$&0.7663&${}^{**}$&-0.4566&${}^{**}$&0.4446\\ 
\\[-2ex]
CR&Winner (W)&-0.0695&&1.0809&${}^{**}$&0.1721&${}^{**}$&0.3739&${}^{**}$&0.1079&${}^{*}$&0.8827\\ 
&Loser (L)&-0.0390&&1.6467&${}^{**}$&0.3710&${}^{**}$&0.5701&${}^{**}$&-0.2617&${}^{**}$&0.8483\\ 
&L -- W&0.0305&&0.5658&${}^{**}$&0.1989&&0.1961&${}^{*}$&-0.3696&${}^{**}$&0.0969\\ 
\\[-2ex]
CMR&Winner (W)&-0.0574&&0.9109&${}^{**}$&0.0860&&0.1779&${}^{**}$&0.1296&${}^{**}$&0.8697\\ 
&Loser (L)&-0.1068&&1.7638&${}^{**}$&0.4647&${}^{**}$&0.7245&${}^{**}$&-0.2785&${}^{**}$&0.8543\\ 
&L -- W&-0.0494&&0.8529&${}^{**}$&0.3787&${}^{**}$&0.5466&${}^{**}$&-0.4081&${}^{**}$&0.3133\\ 
\hline
${}^{**}$ 1\% significance & ${}^{*}$ 5\% significance  
\end{tabular}
}\caption*{The Carhart four-factor regression on the weekly 6/6 contrarian portfolios in U.S. S\&P 500 is given in the table. $\alpha$ is in weekly percentage.}
\label{tbl_carhart_regression_mddr_momentum_weekly_6_6_contrarian_us_spx}
\end{center}
\end{table}

	Comparing with factor exposures of other portfolios, the factor structure of the recovery portfolio is unique. The R portfolio exhibits negative exposures to all the Carhart four factors. Additionally, all the negative factor loadings, except for the exposure on the momentum factor, are statistically significant. Weak dependence on the market factor and the value factor is found in the R portfolio. $R^2$ values of the C, the CR and the CMR portfolios are relatively smaller.
	
	The performance of winner and loser baskets is explicable by the Carhart four-factor model with high $R^2$ values. It is noteworthy that the long (short) baskets from the recovery measures outperform (underperform) the benchmark long (short) basket. Many factor loadings in the basket-level regression are not only positive but also statistically significant. Meanwhile, intercepts of the ranking baskets are negative, except for the long basket in recovery, and not statistically significant in all the cases.
	
\subsection{Monthly momentum strategies}
	Regression intercepts and factor loadings of the alternative momentum strategies are given in Table \ref{tbl_carhart_regression_mddr_momentum_monthly_6_6_momentum_us_spx}. Except for the recovery strategy, all intercepts of the regression model are positive. In particular, the maximum drawdown portfolios outperform the cumulative return strategy in factor-adjusted return. Additionally, the alpha of the M portfolio is not only statistically significant but also the largest one. 	

\begin{table}[h!]
\begin{center}
\caption{Carhart 4-factor regression of monthly 6/6 momentum portfolios in US S\&P 500}
\resizebox{\textwidth}{!}{
\begin{tabular}{l l r @{} l r @{} l r @{} l r @{} l r @{} l r}
\hline
Criterion & Portfolio & \multicolumn{11}{l}{Factor loadings} \\ \cline{3-13} 
 & & \multicolumn{2}{c}{$\alpha(\%)$} & \multicolumn{2}{c}{$\beta_{MKT}$} & \multicolumn{2}{c}{$\beta_{SMB}$} & \multicolumn{2}{c}{$\beta_{HML}$} & \multicolumn{2}{c}{$\beta_{MOM}$} & $R^2$ \\ 
\hline
C&Winner (W)&-0.2004&&0.1465&&-0.3525&${}^{**}$&0.0498&&0.7700&${}^{**}$&0.9068\\ 
&Loser (L)&-0.4174&&2.3585&${}^{**}$&0.8595&${}^{**}$&0.2371&&-0.7340&${}^{**}$&0.8513\\ 
&W - L&0.2169&&-2.2120&${}^{**}$&-1.2120&${}^{**}$&-0.1873&&1.5041&${}^{**}$&0.5962\\ 
\\[-2ex] 
M&Winner (W)&0.0928&&0.4854&${}^{**}$&-0.2204&${}^{**}$&0.0090&&0.1209&${}^{*}$&0.7924\\ 
&Loser (L)&-0.7344&${}^{*}$&2.1554&${}^{**}$&0.8797&${}^{**}$&0.4067&${}^{**}$&-0.4868&${}^{**}$&0.8481\\ 
&W - L&0.8273&${}^{*}$&-1.6700&${}^{**}$&-1.1001&${}^{**}$&-0.3977&${}^{*}$&0.6077&${}^{**}$&0.6468\\ 
\\[-2ex] 
R&Winner (W)&-0.3441&&1.0967&${}^{**}$&0.1019&&0.3911&${}^{**}$&0.1819&${}^{*}$&0.9077\\ 
&Loser (L)&0.0351&&1.2887&${}^{**}$&0.2801&${}^{**}$&0.0096&&-0.1979&${}^{**}$&0.9392\\ 
&W - L&-0.3792&&-0.1920&&-0.1782&&0.3814&${}^{**}$&0.3798&${}^{**}$&0.3809\\ 
\\[-2ex] 
RM&Winner (W)&-0.0857&&0.2903&${}^{**}$&-0.2700&${}^{**}$&0.0356&&0.5195&${}^{**}$&0.9090\\ 
&Loser (L)&-0.5745&&2.2389&${}^{**}$&0.9118&${}^{**}$&0.2944&${}^{*}$&-0.5809&${}^{**}$&0.8583\\ 
&W - L&0.4889&&-1.9485&${}^{**}$&-1.1818&${}^{**}$&-0.2588&&1.1004&${}^{**}$&0.6159\\ 
\\[-2ex] 
CM&Winner (W)&-0.1346&&0.1532&${}^{*}$&-0.3153&${}^{**}$&0.0071&&0.6553&${}^{**}$&0.9037\\ 
&Loser (L)&-0.5945&&2.2985&${}^{**}$&0.9022&${}^{**}$&0.3256&${}^{*}$&-0.6398&${}^{**}$&0.8525\\ 
&W - L&0.4599&&-2.1453&${}^{**}$&-1.2175&${}^{**}$&-0.3185&&1.2951&${}^{**}$&0.6036\\ 
\\[-2ex] 
CR&Winner (W)&-0.2611&&0.3040&${}^{**}$&-0.3229&${}^{**}$&0.1857&${}^{*}$&0.6866&${}^{**}$&0.9061\\ 
&Loser (L)&-0.3838&&2.3090&${}^{**}$&0.8520&${}^{**}$&0.1567&&-0.7319&${}^{**}$&0.8578\\ 
&W - L&0.1226&&-2.0050&${}^{**}$&-1.1749&${}^{**}$&0.0290&&1.4185&${}^{**}$&0.5568\\ 
\\[-2ex] 
CMR&Winner (W)&-0.1384&&0.1791&${}^{*}$&-0.3492&${}^{**}$&0.0306&&0.6891&${}^{**}$&0.9104\\ 
&Loser (L)&-0.5106&&2.3149&${}^{**}$&0.8716&${}^{**}$&0.2500&&-0.6663&${}^{**}$&0.8530\\ 
&W - L&0.3722&&-2.1358&${}^{**}$&-1.2208&${}^{**}$&-0.2194&&1.3553&${}^{**}$&0.5970\\ 
\hline
${}^{**}$ 1\% significance & ${}^{*}$ 5\% significance  
\end{tabular}
}\caption*{The Carhart four-factor regression on the monthly 6/6 momentum portfolios in U.S. S\&P 500 is given in the table. $\alpha$ is in monthly percentage.}
\label{tbl_carhart_regression_mddr_momentum_monthly_6_6_momentum_us_spx}
\end{center}
\end{table}

	Most alternative strategies are exposed to the market, the size and the momentum factors. Different factor structures are found with respect to types of the stock selection rules. The first class of the ranking rules are the selection rules associated with maximum drawdown. Factor-adjusted returns of the maximum drawdown-related portfolios are greater than those of other portfolios including the traditional momentum. Moreover, higher $R^2$ values indicate that larger portions of the performance by the maximum drawdown portfolios are explained by the Carhart four-factor model. 
	
	Contrary to the maximum drawdown-based portfolios, the R portfolio exhibits a smaller intercept than other strategies, and it is the only negative intercept. The recovery portfolio is not significantly related to the market factor and the size factor. Meanwhile, it is dependent on the value factor and the momentum factor. $R^2$ values are relatively lower.
	
	For each portfolio, the performance of the ranking baskets is well-explained by the Carhart four-factor model with high $R^2$ values. It is noteworthy that the long (short) baskets from the maximum drawdown measures outperform (underperform) the benchmark long (short) basket. The differences in intercept and factor loading with respect to ranking criterion are originated from the characteristics in the factor structure of each basket. The winner basket of the recovery momentum strategy exhibits not only a smaller intercept but also large exposure on the market factor than the other winner basket. Meanwhile, the maximum drawdown loser gains the smallest alpha which is the only statistically-significant intercept. The M portfolio is significantly exposed to all the Carhart factors.
	
\section{Conclusion}
\label{sec_mddr_momentum_conclusion}
	In this paper, we introduce alternative ranking criteria using maximum drawdown and recovery, and test their predictability on asset price by implementing monthly momentum and weekly contrarian strategies. Not limited to maximum drawdown and successive recovery, the selection rules include composite indices with cumulative return, maximum drawdown, and recovery. Additionally, the alternative portfolios are tested in South Korea KOSPI 200 universe, SPDR U.S. sector ETFs, and U.S. S\&P 500 universe.
	
	In all the market universes, the alternative selection rules exhibit more robust predictability on asset price by obtaining more persistent profits. For example, the alternative strategies are superior in profitability and volatility to the benchmark strategies. Additionally, the category of the best selection rule in each time scale is consistent in various market universes.
	
	In a weekly scale, the profitability of the contrarian portfolios is improved by using the recovery-related measures, i.e., smaller recovery in the estimation period expects stronger reversion in the future. The R, the CR, and the CMR contrarian portfolios show outperformance over the traditional contrarian strategy. Moreover, the performances are less volatile.
	
	In a monthly scale, maximum drawdown-associated strategies outperform the traditional momentum strategy, i.e., an asset with a smaller maximum drawdown delivers stronger momentum. In general, the CM strategy is the best performer in all the market universes. Additionally, the CMR portfolio and the RM portfolio are as profitable as the CM strategy. Similar to the contrarian cases, the portfolio performances exhibit smaller standard deviations.
	
	Regardless of the time scales, enhanced risk measures of the portfolios indicate that the alternative strategies are less riskier than the benchmark strategies. For example, the alternative portfolios tend to exhibit lower VaR, CVaR, and maximum drawdown in the performance. The similar pattern is also observed at the level of each long/short basket.
	
	The factor analysis also shows that dependence on the stock selection rules exists, and the regression results also support the robust performance of the alternative momentum and contrarian portfolios. In a weekly scale, the factor-adjusted return for the recovery portfolio is not only the largest but also statistically significant intercept. The maximum drawdown portfolio in the monthly horizon achieves a higher intercept than the traditional momentum strategy. Factor exposures are also dependent on ranking rules and types of strategies.
	
\section*{Acknowledgement}
	We are thankful to Young Shin Kim for useful discussions and technical supports on calculation of reward-risk measure by using the ARMA-GARCH model with classical tempered stable innovations.

\end{document}